\documentstyle[multicol,prl,aps,epsfig,floats]{revtex}
\begin{document}
\tightenlines
\title{Interaction-Induced Quantum Dephasing in Mesoscopic Rings}
\author{Dmitri S. Golubev$^{1,4}$, Carlos P. Herrero$^{2}$ 
and Andrei D. Zaikin$^{3,4}$
}
\address{$^1$Institut f\"ur Theoretische Festk\"orperphysik,
Universit\"at Karlsruhe, 76128 Karlsruhe, Germany \\
$^2$Instituto de Ciencia de Materiales de Madrid, CSIC,
Cantoblanco, E-28049 Madrid, Spain \\
$^3$Forschungszentrum Karlsruhe, Institut f\"ur Nanotechnologie,
76021 Karlsruhe, Germany\\
$^4$I.E.Tamm Department of Theoretical Physics, P.N.Lebedev
Physics Institute, 119991 Moscow, Russia
}

\maketitle

\begin{abstract}
Combining nonperturbative techniques with Monte Carlo
simulations we demonstrate that quantum coherence effects for a particle 
on a ring are suppressed beyond a finite length $L_{\varphi}$
even at zero temperature if the particle is coupled to a diffusive
electron gas by means of long range Coulomb interaction. This length
is consistent with $L_{\varphi}$ derived from weak-localization-type of
analysis.
 \end{abstract}


\begin{multicols}{2}


Experiments \cite{MW} strongly suggest that decoherence
time $\tau_{\varphi}$ of electrons in disordered conductors saturates 
to a finite value in the low temperature limit. Theoretical
arguments \cite{GZ1} attribute this saturation to the effect of
electron-electron interactions. This work has initiated intensive
theoretical debates, the present status of which is summarized
in recent publications \cite{debates}.

Although detailed calculations of $\tau_{\varphi}$ for metallic conductors with
disorder and interactions can be rather involved, the main cause
for electron dephasing is quite clear without going into unnecessary
complications: It is the electron interaction with the fluctuating quantum
electromagnetic field produced by other electrons. A question
of fundamental importance here is to understand whether such
interaction can dephase at $T \to 0$, i.e. if the whole interacting
system is in its true {\it quantum mechanical ground state}.   

It is obvious that for a system in thermodynamic equilibrium quantum 
dephasing cannot
be associated with any kind of decay or relaxation in real time, rather
it manifests itself via suppression of off-diagonal elements
of the electron density matrix beyond a certain length $L_{\varphi}$.
Provided there exists nonzero electron dephasing due to its interaction
with quantum environment even at $T\to 0$, this dephasing length  
$L_{\varphi}$
should stay finite down to zero temperature. This, in turn, would imply
that effects sensitive to quantum coherence, such as, e.g., persistent
currents (PC) and Aharonov-Bohm (AB) oscillations in mesoscopic rings,
should be suppressed by interactions at any temperature including $T=0$ if the
perimeter of the ring exceeds $L_{\varphi}$.

Recently the authors \cite{Buttiker} demonstrated that the amplitude of PC can
be reduced by interactions even exactly in the ground state, and this
result was interpreted as a signature of suppression of quantum coherence
at $T=0$. Other authors \cite{Imry} argued that this reduction of PC
is merely a renormalization effect which has nothing to do with dephasing.
Very recently Guinea \cite{Paco} found
that AB oscillations for a quantum particle on a ring interacting
with Caldeira-Leggett bath of oscillators are exponentially reduced beyond
some length $L_{\varphi}$ which is set by interactions and 
remains finite down to zero temperature. Similar result 
was also obtained earlier from a real-time quasiclassical analysis \cite{GZ98}.
It should also be noted that the model \cite{Paco,GZ98} is exactly mapped
onto that of the so-called single electron box, the effective charging energy
of which is well known to reduce exponentially 
at large conductances \cite{many,pa91,WG,HSZ}. 
This reduction immediately translates into 
a finite dephasing length $L_{\varphi}$ at zero temperature found in 
\cite{Paco,GZ98}. 
On the other hand, no finite $L_{\varphi}$ at $T=0$ was found in some 
other models \cite{Paco,Bruder}. 

The purpose of this Letter is to investigate
quantum dephasing for a model with long range Coulomb interactions 
in the presence
of disorder. Our main conclusion is that such interactions can strongly
suppress PC even in thermodynamic equilibrium at $T=0$.

{\it The model and effective action.}
We will consider a quantum particle with mass $M$ on a ring with
radius $R$ threaded by external magnetic flux $\Phi_x$. As in
ref. \cite{Paco} it will be convenient to describe the particle 
position by a vector $\bbox{R}=(R\cos \theta ,R\sin \theta )$
and consider the angle $\theta$ as a quantum variable. We also
assume that the ring is embedded into an effective dissipative environment
and the whole system is in equilibrium at a temperature $T$. 

Our first and standard step is to integrate out the environmental 
degrees of freedom.
After that the grand partition function of the system takes the form
\begin{equation} {\cal Z}=\sum_{m=-\infty}^{\infty}\int_0^{2\pi m}{\cal
D}\theta \exp (i2\pi m\phi_x-S_0[\theta ]-S_{\rm int}[\theta ]).
\label{Z}
\end{equation}
The first term in the exponent takes care of the magnetic
flux while the second term 
\begin{equation}
S_0[\theta ]=\int_0^{\beta}d\tau 
\frac{1}{4E_C}\left(\frac{\partial \theta}{\partial \tau}\right)^2
\label{part}
\end{equation}
is the particle action. Here we defined $\beta=1/T$, $E_C=1/(2MR^2)$ and 
$\phi_x=\Phi_x/\Phi_0$, where $\Phi_0$ is the flux quantum. The term
$S_{\rm int}$ describes the effect of interaction between the particle
and the environment and, hence, depends both on the environment Hamiltonian
and on the form of the interaction. We will assume 
that the particle has the electron charge $e$ which interacts with the
fluctuating electromagnetic field $V$ produced by the effective environment.
For this model one finds
\begin{equation}
S_{\rm int}=-\ln \left\langle \exp \left(-i\int_0^{\beta}d\tau 
eV(\tau , \theta (\tau))\right)\right\rangle_{V}.
\label{eV}
\end{equation}
Provided the fluctuating field $V$ is sufficiently 
well described within the Gaussian approximation, we obtain  
\begin{equation}
S_{\rm int}=\frac{e^2}2\int_0^{\beta}d\tau\int_0^{\beta}d\tau'
 \langle V(\tau , \theta (\tau ))V(\tau' , \theta (\tau' )) \rangle .
\label{Sint}
\end{equation}
The correlator $\langle VV\rangle$ can be expressed via the dielectric
susceptibility of the environment $\epsilon (\omega , k)$ as:
\begin{equation}
\langle VV \rangle =
T\sum_{\omega_n}\int\frac{d^3k}{(2\pi)^3}\frac{4\pi}{k^2\epsilon
(i|\omega_n|,k)}e^{-i\omega (\tau -\tau')+
i\bbox{kX}},
\label{VV}
\end{equation}
where $\omega_n=2\pi nT$ and $\bbox{X}=\bbox{R}(\tau
)-\bbox{R}(\tau')$. In what follows we will model the 
environment by a 3d diffusive electron gas with 
$1/\epsilon (\omega , k)\approx (-i\omega +Dk^2)/(4\pi \sigma)$, where $\sigma$
is the Drude conductivity of this gas and $D=v_Fl/3$ is the electron
diffusion coefficient. Substituting $1/\epsilon (i|\omega |, k)$ into 
(\ref{VV}) and performing integrations we observe that
the term $\sim Dk^2$ yields only an $X$-independent energy shift. 
Integrating the remaining 
contribution $\sim |\omega_n|$ over $k \lesssim 1/l$ in (\ref{VV}) 
we obtain the result $\propto
{\rm min} (1/X,1/l) \approx 1/\sqrt{X^2+l^2}$. Defining 
$X=2R\sin [(\theta(\tau )-\theta (\tau'))/2]$ and summing over
$\omega_n$, from (\ref{Sint}) and (\ref{VV}) we get
\begin{eqnarray}
S_{\rm int}[\theta ]=\alpha\int_0^{\beta}d\tau
\int_0^{\beta}d\tau'\frac{\pi^2T^2K(\theta (\tau )-\theta (\tau'))}
{\sin^2 [\pi T(\tau -\tau')]},
\label{actionD}\\
K(z)=1-\frac{1}{\sqrt{4r^2\sin^2(z/2)+1}}, 
\label{F}
\end{eqnarray}
where $\alpha =3/(8k_F^2l^2)$ and $r=R/l$. 
The integral in eq. (\ref{actionD}) is understood
as a principal value. The divergence at $\tau=\tau'$ is
then regularized in a standard manner by requiring  $K(0)=0$ which 
explains the origin of the first term in (\ref{F}).
Below we will set $1/k_F \ll l \ll R$ implying 
that interaction is effectively weak $\alpha \ll 1$ while the 
parameter $r=R/l$ is large $r\gg 1$. 

{\it Perturbation theory.}
In the non-interacting limit $S_{\rm int} \to 0$  
the partition function (\ref{Z}) is trivially evaluated, and
for the flux-dependent part of the ground state energy at 
$-0.5 <\phi_x \leq 0.5$ one finds $E_0(\phi_x)=E_C\phi_x^2$.

Let us now take interaction into account. We first assume that interaction
effects are weak, in which case it suffices to expand 
the partition function (\ref{Z}) to the first order in 
$S_{\rm int}$ (\ref{actionD}).
Here it will be convenient for us to rewrite the function $K$ 
in terms of the Fourier series \cite{Paco} 
\begin{equation}
K=\sum_{n}a_n\sin^2\left[\frac{n(\theta (\tau)-
\theta (\tau'))}2\right],
\label{Fourier}
\end{equation}
where $a_n \sim (2/\pi r)\ln (r/n)$ for $1\leq n \lesssim r$ and
$a_n \approx 0$ otherwise. After that the
calculation reduces to a simple Gaussian integration 
over the $\theta$-variable in each term of the series. 
Performing this integration and summing
over $n$ one obtains $E_0(\phi_x)$ and finds a {\it diamagnetic} current 
$I = (e/2\pi)(d E_0/d \phi_x)$ with the result
\begin{equation}
I= \frac{eE_C}{\pi}\left[\phi_x-\frac{\alpha}{2}\sum_{n=1}^rna_n
\ln \left(\frac{n+2\phi_x}{n-2\phi_x}\right)\right],
\label{pert}
\end{equation}
where $-0.5 <\phi_x \leq 0.5$. The last term in (\ref{pert}) represents 
the first order correction
to PC in the ring due to Coulomb interaction at $T=0$. 
This correction is negative, i.e. 
for the problem in question interactions {\it suppress} 
PC even {\it in the ground state}.
For $\phi_x \ll 1$ the interaction term in (\ref{pert}) is
linear in $\phi_x$ and it is small in the parameter
$\alpha \sum_{n=1}^ra_n \sim \alpha  \ll 1$
as compared to the non-interacting PC. A somewhat stronger effect of 
interactions $\sim \alpha \ln r$ was found in ref. \cite{Paco}
within the renormalization 
group (RG) analysis. Below we will show that  
already for small $\alpha \ll 1$ suppression 
of quantum coherence is actually {\it much stronger} than it could be 
expected both from eq. (\ref{pert}) and the RG approach \cite{Paco}.

{\it Nonperturbative effects.} Let us first consider
the regime of not very low temperatures. In this case the partition
function (\ref{Z}) can be evaluated semiclassically. As in 
\cite{WG,HSZ} one can find the classical paths providing
the minimum of the total action $S=S_0+S_{\rm int}$ for any 
winding number $m$. These are the so-called ``straight line'' paths
$\theta_{\rm cl} (\tau )= 2\pi mT\tau$,
describing rotations of the particle around the ring. Substituting
the paths $\theta_{\rm cl} (\tau )$ into the sum (\ref{Z}) one gets 
\begin{equation}
{\cal Z} \sim \sum_{m=-\infty}^{\infty}\exp \left(i2\pi m\phi_x-\frac{\pi^2m^2T}{E_C}-4\pi |m|\alpha r\right).
\label{scl}
\end{equation}
In the limit $T \gg E_C/\pi^2$ it suffices to keep only the terms 
with $m=0,\pm 1$ and the flux-dependent part of the free energy
$F=-T\ln {\cal Z}$ can easily be evaluated. Then for PC $I(T) = (e/2\pi)(d F/d
\phi_x)$ one obtains
\begin{equation}
I=2eT\exp \left(-\frac{\pi^2T}{E_C}-4\pi 
\alpha r\right)\sin (2\pi \phi_x).
\label{highT}
\end{equation}
We observe that in addition to trivial reduction of PC due to thermal
fluctuations there is also a $T$-independent term which yields 
exponential suppression of $I$ provided
the perimeter of the ring $2\pi R$ exceeds the value
\begin{equation}
L_{\varphi} \sim l/\alpha \sim l (k_Fl)^2.
\label{dephl}
\end{equation}
Eq. (\ref{dephl}) defines the dephasing length for the problem 
in question. This length does not depend on temperature and is 
controlled by the effective interaction strength $\alpha$.  

The above semiclassical analysis 
is justified at sufficiently high temperatures, whereas at lower
$T \lesssim E_C$ fluctuations around the classical paths
$\theta_{\rm cl} (\tau )$ become important and our treatment needs 
to be modified. 
Provided the suppression of PC is sufficiently strong one can
employ the instanton technique \cite{pa91,WG}.

Nontrivial saddle points $\tilde \theta (\tau )$
which describe quantum tunneling of the
variable $\theta$ between different topological sectors, e.g. between 
the states $\theta =0$ and $\theta =2\pi$, are defined as solutions
of the equation $\delta S/\delta \tilde \theta =0$.
For the action (\ref{actionD}), (\ref{F}) this equation is
approximately satisfied for a wide class of sufficiently smooth functions
obeying the boundary conditions $\tilde \theta (0)=0$ and $\tilde
\theta (\beta )=2\pi$. Defining the typical instanton frequency
$\Omega$, one finds \cite{pa91,WG} that
for $\Omega \lesssim E_C$ and $T \ll E_C$
the total action $S[\tilde \theta (\tau )]$ does not depend on $E_C$ 
and is equal to $S_{\rm int}[\tilde \theta (\tau )]=4\pi \alpha r$, while
for  $\Omega > E_C$ the instanton contribution gets suppressed by
the kinetic energy term $S_0$. For 
$S_{\rm int}[\tilde \theta (\tau )] \gg 1$  
both the tunneling amplitude between the
states  $\theta =2\pi m$ and $\theta =2\pi (m\pm 1)$ and the flux-dependent
part of the free energy $F$ are exponentially
reduced as $\propto \exp (-S[\tilde \theta (\tau )])$. Hence,
in the limit of small $\phi_x \ll 1$ we obtain $I=eE_C^*\phi_x/\pi$, where 
\begin{equation}
E_C^*/E_C =A(T) \exp (-4\pi\alpha r).
\label{ren}
\end{equation}
This result is justified provided $4\pi \alpha r \gg 1$ and the ratio
$E_C^*/E_C$ is sufficiently small. It demonstrates that also at low $T \ll E_C$
PC is strongly suppressed by Coulomb
interaction provided the ring perimeter $2\pi R$ exceeds 
the dephasing length $L_{\varphi}$ (\ref{dephl}). 
Eq. (\ref{ren}) suggests that at low $T$ 
interaction-induced decoherence in our model is controlled by the 
parameter $\alpha r \sim \alpha \sum_{n=1}^rna_n$
rather than by $\alpha$ or $\alpha \ln r$. Hence, for $r \gg 1$ 
the decoherence effect of Coulomb interactions is {\it much stronger}
than it could be expected both from perturbative and RG approaches.

While the above instanton analysis is
applicable for $4\pi \alpha
r \gg 1$, it can hardly provide a quantitative description of PC
for moderate $\alpha r$. Furthermore, even for large rings rigorous analytic
evaluation of the pre-exponent $A(T)$ in eq. (\ref{ren}) is difficult 
because the action (\ref{actionD}), (\ref{F}) is
strongly non-Gaussian and the standard techniques are hard to employ. On 
a qualitative level one expects that many non-Gaussian
quasi-zero modes should
yield a large pre-exponent $A(T)$ {\it increasing} upon
decreasing $T$.

{\it Quantum Monte Carlo analysis.} In order to quantitatively
describe reduction of PC at arbitrary values of $\alpha r$
we carried out extensive Monte Carlo (MC) simulations and
directly computed $E_C^*$ as a function of $\alpha$ and $r$.
Our method is described in details in ref. \cite{HSZ}. In order to
discretize the action defined by eqs. (\ref{part}), (\ref{actionD}) and 
(\ref{F}) we introduce the Trotter number $N=\beta E_C$. 
We then evaluate path integrals in eq. (\ref{Z}) for different $m$ 
and determine $E_C^*(T)=2\pi^2T\langle m^2\rangle_{\phi_x=0}$. By increasing
$N$ we decrease effective temperature in our simulations. The
value $E_C^*(T)$ grows with decreasing $T$ and finally reaches a plateau
which defines the zero temperature value  $E_C^* (\alpha ,r)$, 
see the inset in Fig. 1.
\vskip  -1.5 truecm
\begin{center}
\leavevmode \epsfxsize=3in
\epsfbox{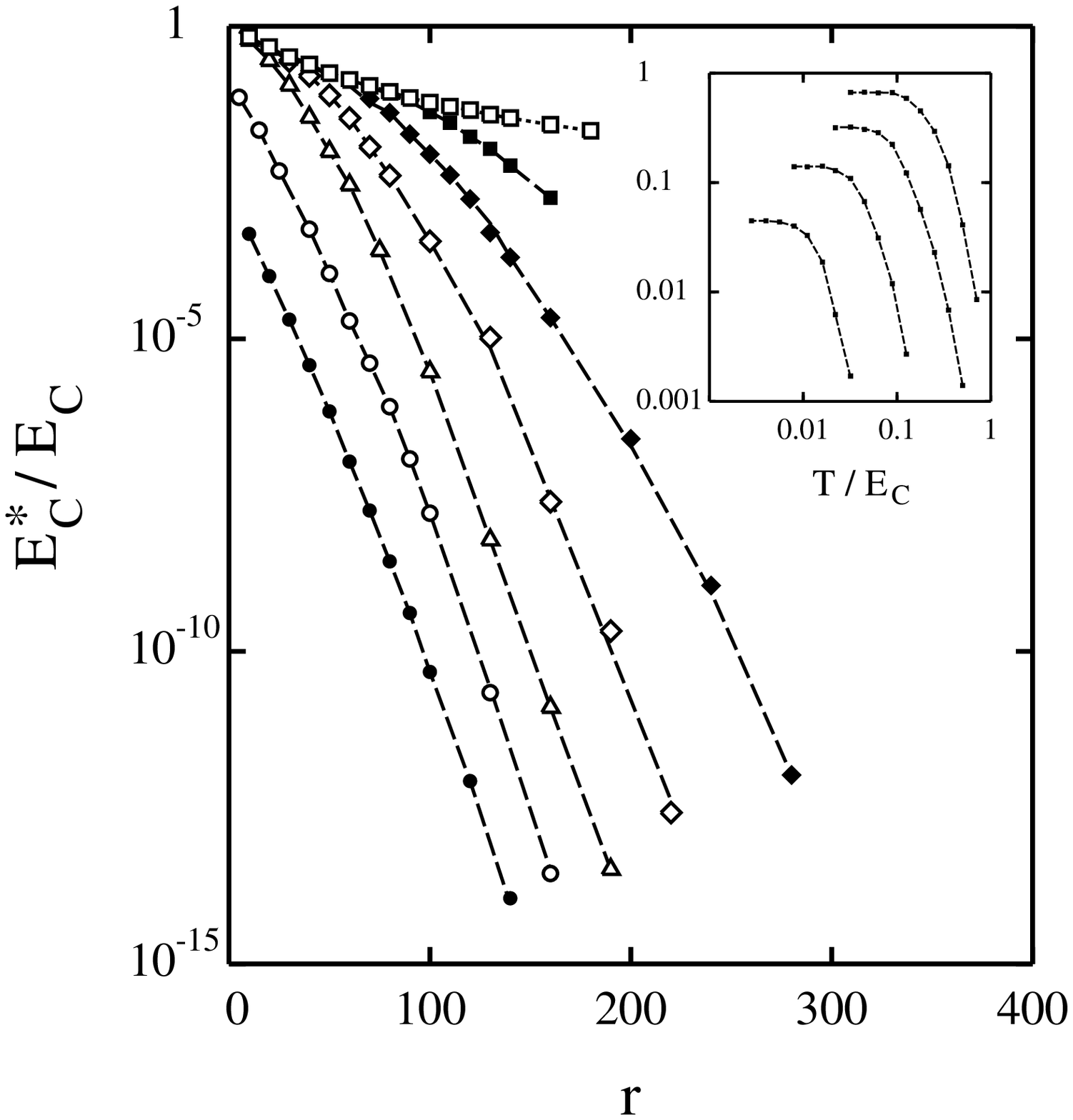}
\end{center}
\vskip  -0.25 truecm
\begin{small}
{\bf Fig. 1} The value $E_C^*/E_C$ as a function of $r$ for 
$\alpha = 0.019$ and different $T$. From bottom to top: 
$T/E_C$ = 1, 0.5, 0.14, 0.06, 0.03 and 0. 
The inset: $E_C^*/E_C$ as a function of $T$. From top to bottom: $r$ =
10, 30, 60, and 120.\\
\end{small}
At relatively high temperatures
$T \gtrsim \alpha rE_C$ our MC data confirm the semiclassical
result (\ref{highT}).
Also at lower $T$ we observe dramatic suppression of $E_C^*$
with increasing $r$, see Fig. 1. At sufficiently large $r$ this suppression
turns exponential and at $E_C / (4 \pi \alpha r) \lesssim T \lesssim E_C$ is well described by eq. (\ref{ren}) with 
$A(T) \approx\exp (cE_C/T)$ and $c \sim 1$.
Thus, eq. (\ref{ren}) holds even  
at temperatures {\it well below} the level spacing ($\sim E_C$) of
the noninteracting problem. 
\vskip  -1.5 truecm
\begin{center}
\leavevmode \epsfxsize=3in
\epsfbox{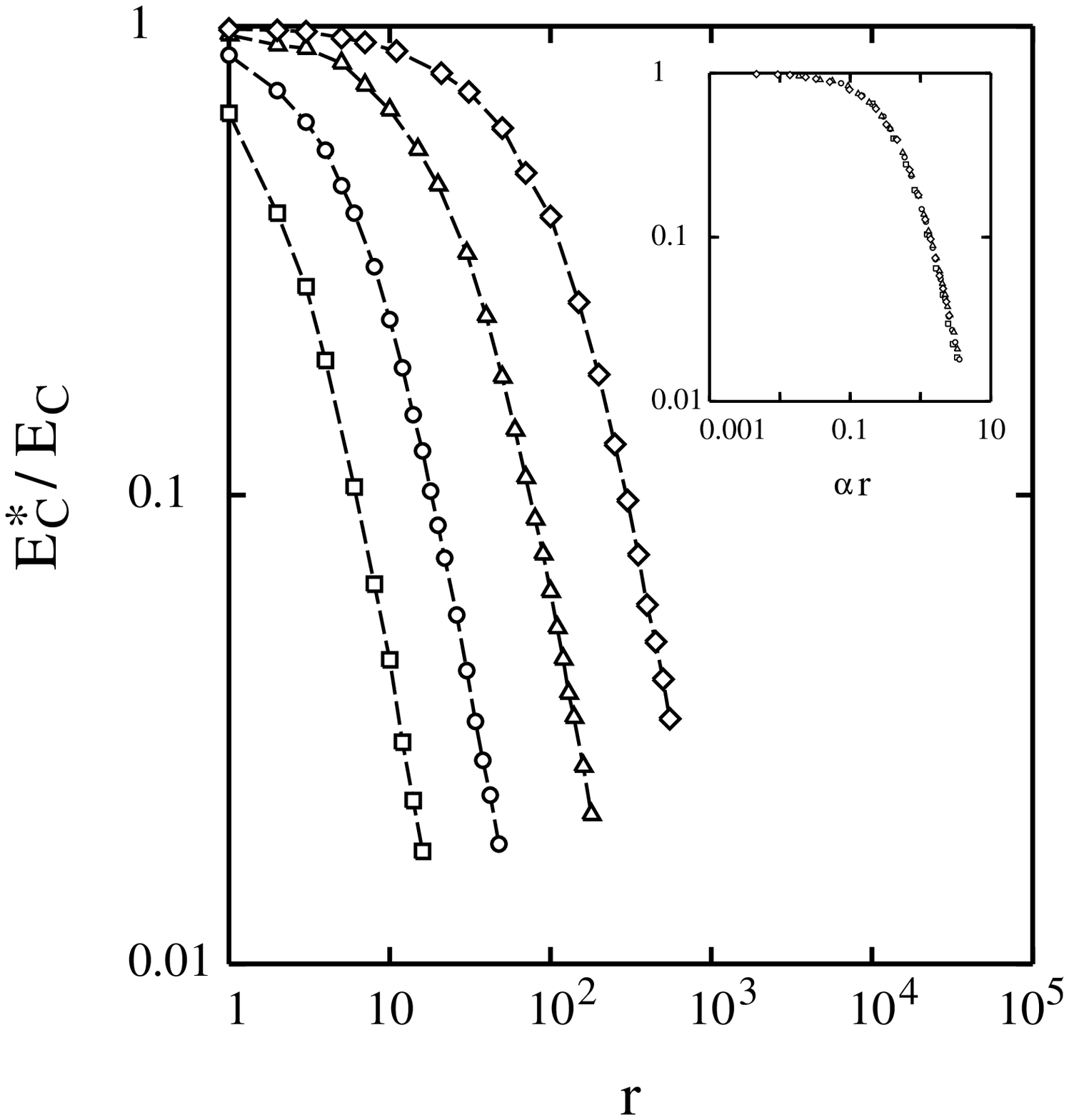}
\end{center}
\vskip  -0.25 truecm
\begin{small}
{\bf Fig. 2}
$E_C^*/E_C$ as a function of $r$ at $T=0$ (bottom to top: $\alpha$ = 0.21, 0.07, 0.019, and
0.005). The inset: the same data for different $\alpha$ collapse onto 
one curve if plotted versus $\alpha r$.\\
\end{small}
\vskip -0.5 truecm 
The data corresponding to the exact limit $T=0$ are displayed in Fig. 2 for
different $\alpha$. At relatively small $r$ we observe
very weak suppression of $E_C^*$ in agreement with our perturbative
result (\ref{pert}). However, at $\alpha r \gtrsim 1$ the ratio
$E_C^*/E_C$ rapidly decreases as $E_C^*/E_C\propto r^{-\gamma}$
with $\gamma \approx 1.8$. In order to judge whether at $T=0$ such power law
behavior holds at all $r\gtrsim 1/\alpha$ or it is merely
an intermediate regime with further crossover to the exponential dependence
(\ref{ren}) (cf. Fig. 1) it is necessary to extend our simulations at large $N$
to values $E_C^*/E_C$ below 0.01. Such details are, 
however, of no significance for our conclusions.
Much more importantly, our MC data clearly demonstrate drastic
suppression of PC by Coulomb interactions even exactly at $T=0$ 
in rings with $2\pi R > L_{\varphi}$ as well as the failure of perturbation
theory for such values of $R$. Hence,  $L_{\varphi}$ (\ref{dephl}) 
remains the relevant scale down to $T=0$, see the inset in Fig. 2.

{\it Real time calculation.}
In order to illustrate a relation between the above calculation and 
real time quasiclassical analysis let us evaluate the 
return probability for a pair
of classical paths which encircle the ring and return to the 
initial point after time $t$. 
As in \cite{GZ98} we distinguish two probabilities $W_1$ and $W_2$ 
corresponding
respectively to two identical and two time-reversed paths. The
quantity $W_1$ is the standard return probability and $W_2$ describes
quantum interference between time-reversed paths. The latter quantity
is sensitive to quantum coherence and it vanishes
in the classical limit. 

Without interactions the
time reversal symmetry yields  $W_1=W_2=W^{(0)}$. The effect
of interactions is accounted for by the influence functional \cite{FH}
${\cal F}=\exp (iS_R-S_I)$, and the probabilities are now
equal to $W^{(0)}{\cal F}$. For our model the actions $S_R$ and
$S_I$ are evaluated by integrating out the fluctuating electromagnetic
field produced by the electron gas. Similarly to eqs. (\ref{Sint}),
(\ref{VV}) both
$S_R$ and $S_I$ are expressed in terms of the inverse dielectric
function $1/\epsilon (\omega ,k)$ of the electron environment.  

For two identical paths one finds $S_R=S_I=0$, i.e. 
the probability $W_1=W^{(0)}$ is not
affected by interactions at all. For a pair of time-reversed paths one also has
$S_R=0$, but the action $S_I$ is now positive $S_I >0$ and,
hence, $W_2=W^{(0)}\exp (-S_I)$. 
In order to evaluate $S_I$ it is convenient to ``unfold'' the
time-reversed path into two straight lines 
$\bbox{r}_1(t_1)=\bbox{v}t_1$ and $\bbox{r}_2(t_1)=\bbox{v}(t-t_1)$,
where $0\leq t_1\leq t$.
Inserting these paths into $S_I$  \cite{GZ1} we get
\begin{eqnarray}
S_I&=&\frac{e^2}{2}\int\frac{d\omega}{2\pi}\int\frac{d^3k}{(2\pi)^3}\,
{\rm Im}\left(\frac{-4\pi}{k^2\epsilon(\omega,k)}\right)\coth\frac{\omega}{2T}
\nonumber\\ &&\times\,
\int_0^t dt_1 \int_0^t dt_2\,{\rm e}^{-i\omega(t_1-t_2)}{\cal I}(t_1,t_2),
\end{eqnarray}
where ${\cal I}=2\cos (\bbox{kv}t_-)-2\cos [\bbox{kv}(t-t_+)]$ and $t_{\pm}=t_1\pm t_2$. In the long time limit one finds
\begin{eqnarray}
S_I
&=&  \frac{e^2t}{2\pi v}\int_0^{1/l}dk\int_{-kv}^{kv}d\omega\; \frac{\omega\coth\frac{\omega}{2T}}{4\pi\sigma k}. 
\end{eqnarray}
Defining the path length $L=vt$, at $T=0$ we obtain
\begin{equation}
S_I=\frac{e^2 L}{16\pi^2\sigma l^2}\sim\frac{L}{l\alpha}.
\label{rt}
\end{equation}
Hence, similarly to PC, at $T\to 0$ the probability $W_2$ is
suppressed by interactions beyond the length (\ref{dephl}) 
and the ratio $W_2/W_1 \propto \exp (-L/L_{\varphi})$ 
vanishes at large $L$. The latter result
proves that reduction of PC is due to zero temperature decoherence and 
is {\it not} just a renormalization effect. The same conclusion follows from
studying the AB conductance which we do not present here.

Finally, let us define the dephasing time $\tau_{\varphi}$. For ballistic
particle motion considered here we set
$\tau_{\varphi}=L_{\varphi}/v$. In particular, for $v\sim v_F$ 
we get $\tau_{\varphi}\sim \tau_e(k_Fl)^2$,
where $\tau_e=l/v_F$ is the elastic scattering time for electrons. The latter
expression for $\tau_{\varphi}$ is identical to the electron dephasing
time obtained in ref. \cite{GZ1} for 3d disordered conductors. This
observation is consistent with the idea \cite{M} that the
results of weak localization and persistent current experiments may be
closely related. In particular, we
expect both to yield the same dephasing length $L_{\varphi}$ at low $T$.

We are grateful to F. Guinea, M. Paalanen and G. Sch\"on for interesting
discussions.

\end{multicols}

\end{document}